\documentclass{sig-alternate-per-modified}
\usepackage{tikz}
\usepackage{wrapfig}
\usepackage{booktabs}
\usepackage{amsmath}
\usepackage{amsfonts}
\usepackage{xspace}
\usepackage{xcolor}
\usepackage[noend]{algorithm2e}

\newcommand{\mtc}{\tau _c}  
\newtheorem{definition}{Definition}
\newtheorem{proposition}{Proposition}

\newtheorem{lemma}{Lemma}
\usepackage{subfigure}
\usepackage{enumitem}
\usepackage{wrapfig}
\usepackage{soul}
\usepackage{hyperref}
\begin{document}
\title{TTL Approximations for a Family of CLOCK-Like Cache Replacement Policies}

\numberofauthors{3}
\author{
\alignauthor
Yirong Wang\\
       \affaddr{Northeastern University}\\
       \affaddr{Boston, MA, USA}
\alignauthor
Peter Desnoyers\\
       \affaddr{Northeastern University}\\
       \affaddr{Boston, MA, USA}
\alignauthor
Benny Van Houdt\\
       \affaddr{University of Antwerp}\\
       \affaddr{Antwerp, Belgium}
}
\maketitle

\begin{abstract}
  While CLOCK-like cache replacement policies are highly appealing in practice due to their frugal memory usage and implementation simplicity, analytical models for their performance remain limited.
  The difficulty stems from the "eviction hand" itself. As it scans sequentially for an evictee, sparing objects with survival counters along the way, it inherently entangles the state of one cached object with the next. This strong coupling heavily obscures the system's steady state.
  
  To address this gap, we introduce a time-to-live (TTL) approximation for a family of eviction policies---which we term CLOCK($K$)---that decouples individual object dynamics using a globally fixed timer. In contrast to the intrinsic coupling induced by assuming a fixed cache capacity, such a TTL proxy yields a tractable matrix-analytic solution for the steady-state miss ratio functions across a broad and highly expressive class of phase-type renewal traffic. Among other benefits, this effectively reduces the analysis per-object to solving a linear system of order $O(Kd)$, where $d$ is the phase dimension. Moreover, our model for CLOCK($K$) recovers the canonical CLOCK when $K=1$ and FIFO when $K=0$, while empirically exhibiting Schur-concavity under the Independent Reference Model (IRM). Finally, we validate the model against a suite of trace-driven simulations. 

\end{abstract}

\section{Introduction} 

Recently, CLOCK-like policies have re-emerged as lightweight yet performant cache-replacement designs, owing to their small state and low update overhead.
Despite their practical appeal, analytical models for their performance remain scarce. 
In particular, while FIFO already admits TTL-style approximations~\cite{dan_approximate_1990, garetto2016unified}, comparably tractable miss-ratio models for CLOCK-like policies are limited to more amenable variants. For instance, clock-per-request (CpR)~\cite{hasslinger_scope_2023}, which advances the clock hand on every request rather than on eviction, admits the same IRM product-form solution as FIFO and RANDOM. 
Alternatively, Ran-CLOCK~\cite{ran-clock} replaces the deterministic scan-based eviction path with random probing, thereby producing a mean-field fixed point. Despite progress on these variants, no model for classical CLOCK is found in the literature.

We address this gap by analyzing a family of eviction policies, termed CLOCK($K$)~\cite{corbato_paging_1968, ran-clock} in which FIFO corresponds to $K=0$, the canonical CLOCK corresponds to $K=1$, and larger $K$ increasingly utilizes access frequency while preserving a parsimonious circular-array implementation. 
Concretely, we predict miss ratio curves for CLOCK($K$) under a TTL approximation~\cite{fagin1977asymptotic, hao_che_hierarchical_2002}, in the sense that a cache with a capacity constraint is replaced by a TTL queue with a fixed residence time. The per-object local dynamics from admission to eviction comprises multiple traversals of such a queue, each of which corresponds to a full rotation of the eviction hand, during which an object may be hit multiple times and have its \emph{access counter} incremented up to $K$. 
The analytical challenge stems from this recycling mechanism: the eviction hand does not simply purge the oldest item, but rather inspects and resets access counters, allowing recently accessed objects to survive and cycle back through the buffer. Consequently, the eviction of an object is strongly affected by the counter states of its neighbors visited immediately before it.

Our approach decouples object states by assuming a globally fixed TTL that is uniquely determined by the cache-capacity constraint, chosen such that the expected object-occupancy equals the target cache size. Thus the global coupling of the eviction hand is compressed into one capacity equation, whereby the local object dynamics become finite-dimensional and tractable.
For the broad family of phase-type (PH) renewal traffic, this approach yields a clean matrix-analytic model. The in-traversal evolution can be described by a continuous-time Markov chain (CTMC) over the joint PH phase and access counter state. Sampling this chain at traversal boundaries, where the eviction-hand rule applies, gives an absorbing discrete-time Markov chain (DTMC) in which absorption corresponds to eviction. The fundamental matrix of this DTMC directly yields per-object cache occupancy and miss ratio. Thus, computing the per-object miss ratio curve reduces to solving a linear system of dimension $O(Kd)$, where $d$ is the PH order.

Beyond analytical predictions, we {empirically demonstrate} that under IRM, CLOCK($K$) exhibits Schur-concavity with respect to object popularity.


\section{Preliminaries}
\label{sec:background}
Consider a cache of capacity $C$ storing objects from a fixed catalog $\{1,\dots,M\}$ with $M>C$.
All objects have equal size, so the cache can hold up to $C$ objects, never more.
A cache replacement algorithm is used to determine which object to evict on a cache miss when the cache is full.
\subsection{Cache Replacement Algorithms}
\label{sec:replacement_algorithms}
\begin{figure}
  \centering
  \includegraphics[width=1\columnwidth]{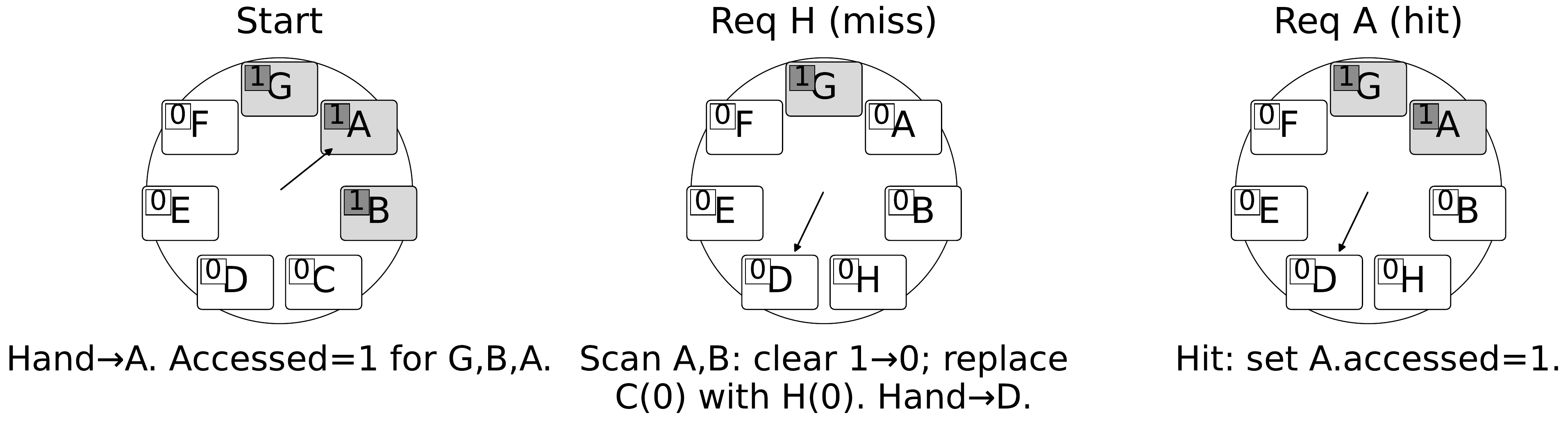}
  \caption{"Rotating hand" representation of CLOCK. Each panel shows the cache state immediately after processing the corresponding request.}
  \label{fig:clock_a}
\end{figure}

\begin{figure}
  \centering
  \includegraphics[width=1\columnwidth]{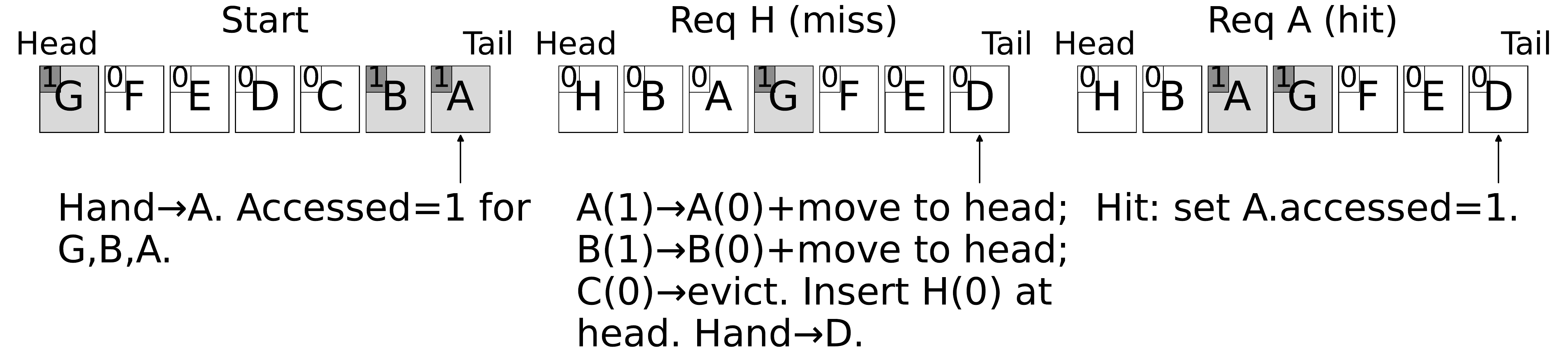}
  \caption{"FIFO with second chance" representation of CLOCK. Each panel shows the cache state immediately after processing the corresponding request.}
  \label{fig:clock_b}
\end{figure}

\noindent\textbf{CLOCK.} It is typically visualized as a circle of objects with a ``rotating hand'' making eviction decisions. Fig.~\ref{fig:clock_a} illustrates this representation. Each cached object carries an \emph{accessed bit} ${A} \in \{0,1\}$, which is set to $1$ on every access (hit). When a cache miss occurs and a new object must be inserted, the eviction hand scans the objects clock-wise until it finds an object with ${A} = 0$; this object is evicted and replaced with the new object, whose ${A}$ bit is initialized to $0$. If the eviction hand encounters an object with ${A} = 1$, it clears the bit to $0$ and continues scanning.

While several variants of CLOCK have been described, we find that a representation, namely "FIFO with second chance", is more intuitive for a TTL-based analysis. This illustration is shown in Fig.~\ref{fig:clock_b}.
Here we view the CLOCK cache as a FIFO queue wherein each object with its access bit set receives a "second chance" upon tail inspection. Both new items (i.e., on cache misses) and recycled items enter at the head; their ${A}$ bit is cleared on insertion, and will be set if the object is accessed while traversing the queue.

\noindent \textbf{CLOCK($K$).} 
This family generalizes the \emph{access bit} of CLOCK to an \emph{access counter} $k \in \{0, 1, \cdots, K\}$: on a miss, the eviction hand scans sequentially, decrementing any positive counter by one, advancing until it finds a victim with $k=0$ to replace. New objects are inserted at the current hand position with $k = 0$. 

Under an equivalent "FIFO with multiple chances" representation, this dynamic translates to a deterministic rule applied at the queue's tail: objects with $k=0$ are evicted, whereas recycled objects are routed back to the queue's head with $k$ decremented by one.

With $K{=}1$, CLOCK($K$) is the classic CLOCK; with $K{=}0$ CLOCK($K$) degenerates to FIFO.

\subsection{Traffic Models}
\label{sec:traffic_models}
We model the request stream as the superposition of $M$ independent object-level processes, one per object $i\in\{1,\dots,M\}$. We normalize the total request rate to $1$, so object $i$ has long-term request rate $\lambda_i$ with $\sum_{i=1}^{M}\lambda_i=1$. Under this normalization, $\lambda_i$ is also the popularity weight $p_i$.

We assume time is continuous, and requests arrive according to one of two workload models:
\subsubsection{Independent Reference Model}
\label{sec:irm}
Each object $i$ is requested by an independent Poisson process of rate $\lambda_i$. The superposition has a unit arrival rate under the normalization $\sum_{i=1}^{M}\lambda_i=1$.



\subsubsection{Phase-Type Renewal Model}
\label{sec:ph}
Each object $i$ has inter-arrival time
\[
A_i \sim \mathrm{PH}(\boldsymbol{\alpha}_i^{\top},\mathbf T_i),
\]
i.e., the time to absorption of a finite-state CTMC with initial phase distribution
$\boldsymbol{\alpha}_i^{\top}$ and transient subgenerator $\mathbf T_i$. Let $\mathbf 1$
denote the all-ones column vector, the absorption-rate vector is $\boldsymbol{\mu}_i = -\mathbf T_i \mathbf 1$.
The mean inter-arrival time is
\[
\mathbb E[A_i] = -\boldsymbol{\alpha}_i^{\top} \mathbf T_i^{-1}\mathbf 1,
\]
and hence the request rate is
\[
\lambda_i =
\frac{1}{-\boldsymbol{\alpha}_i^{\top} \mathbf T_i^{-1}\mathbf 1},
\qquad
\sum_{i=1}^{M}\lambda_i = 1.
\]

\section{Analytical Model of CLOCK($K$)}
\label{sec:analysis}

\begin{figure}[h!]
  \centering
  \includegraphics[width=1\columnwidth]{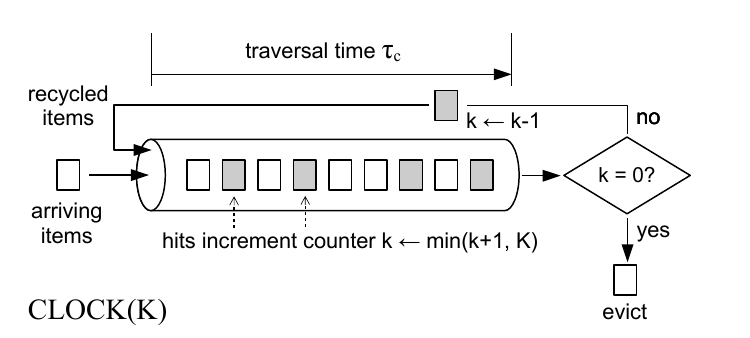}
  \caption{The TTL-queue approximation for CLOCK($K$). Objects undergo deterministic traversals of length $\mtc$ from head to tail, during which hits increment their counter $k$ up to $K$. Upon reaching the tail, objects with $k=0$ are evicted; otherwise, their counters are decremented by one and recycled to the head for next traversal.}
  \label{fig:clockk_queue}
\end{figure}

We develop a predictive analytical framework to characterize the steady-state miss ratio of CLOCK($K$). Our approach relies on a TTL-based queue representation. Here, objects undergo deterministic queue-\emph{traversals} labeled $\ell = 0,1,\dots$, each of fixed duration $\mtc\in\mathbb R_+$; see Figure~\ref{fig:clockk_queue}. 
Upon admission, an object joins the head of the queue. A traversal completes when the object reaches the tail eviction hand, taking exactly real time $\mtc$. Hits during this window do not alter the object's queue position. Instead, they increment its access counter $k$ which is capped at $K$. At the tail, a strict deterministic rule applies: if $k = 0$, the object is evicted; if $k > 0$, it is recycled to the head for traversal $\ell+1$, and its counter drops by one.

To characterize a cache under this approximation, we first consider an individual object $i$. Let $H_i(\mtc)$ denote the expected total hits the object accrues across all traversals from admission to eviction. Because an object is admitted strictly upon a miss, its cache residence consists of exactly one miss followed by an expected $H_i(\mtc)$ hits. The per-object miss probability thus evaluates natively to the ratio of misses to total accesses:
\begin{equation}
P_{\mathrm{miss},i}(\mtc) = \frac{1}{1 + H_i(\mtc)}.
\label{eq:pmiss_i}
\end{equation}
The aggregate miss ratio under the request rate mix $\{\lambda_i\}_{i=1}^M$ is then the weighted sum:
\begin{equation}
P_{\mathrm{miss}}(\mtc)=\sum_{i=1}^{M}\lambda_i\,P_{\mathrm{miss},i}(\mtc).
\label{eq:pmiss_agg}
\end{equation}
{Let} $V_i(\mtc)$ {denote} the expected number of traversals object $i$ completes before eviction. Its expected residence time is simply $\mtc\,V_i(\mtc)$. Again since an object enters the cache strictly upon a miss, its admission rate is $\lambda_i P_{\mathrm{miss},i}(\mtc)$. By Little's law, the expected number of cached copies of object $i$ is precisely 
\begin{equation}
\mathcal C_i(\mtc) = \mtc \lambda_i P_{\mathrm{miss},i}(\mtc) V_i(\mtc).
\label{eq:occ_i}
\end{equation}
Thus the global capacity constraint follows as
\begin{equation}
\mathcal C(\mtc) = \mtc \sum_{i=1}^M \lambda_i\,P_{\mathrm{miss},i}(\mtc)\,V_i(\mtc).
\label{eq:tc_clockK}
\end{equation}
Note that~\eqref{eq:tc_clockK} naturally reduces to FIFO when $K=0$, wherein every admission makes exactly one traversal ($V_i(\mtc)=1$ for all $i$).

The analytical challenge is to determine $V_i(\mtc)$ and $H_i(\mtc)$ for a given $\mtc$. In the following, we demonstrate one such solution for PH arrivals described in \S\ref{sec:ph}, which subsumes IRM (\S\ref{sec:irm}) as a special case.

\subsection{Phase-Type Renewal Model}
\label{sec:clockK_ph}
For clarity of exposition, we suppress the object index in analysis and restore it at the end. Thus consider a single object with inter-request time distribution $\mathrm{PH}(\boldsymbol{\alpha}^\top,\mathbf T)$ of order $d$. Within a traversal of duration $\mtc$, the joint evolution of the PH phase and the counter forms a CTMC on the state $(j,k)$. Here $j\in\{1,\dots,d\}$ is the phase and $k\in\{0,1,\dots,K\}$ is the counter. Grouping states by counter yields a state space of dimension $(K{+}1)d$:
\[
(1,0),\dots,(d,0),\ (1,1),\dots,(d,1),\ \dots,\ (1,K),\dots,(d,K).
\]
Let $\mathbf{x}^{(\ell)}$ be the state distribution vector at the head of traversal $\ell$. A new admission enters at counter $0$ with phase distributed as $\boldsymbol{\alpha}^\top$:
\begin{equation}
\mathbf{x}^{(0)} = (\boldsymbol{\alpha}^\top,\mathbf{0},\dots,\mathbf{0}).
\label{eq:ph_init}
\end{equation}
A generator $\mathbf{A}\in\mathbb{R}^{(K+1)d\times (K+1)d}$ must govern the in-traversal CTMC. To construct this generator, let $\boldsymbol{\mu}:=-\mathbf{T}\mathbf{1}$ be the phase-dependent arrival rates. Now, note that transitions fall into two categories: i) phase changes occur without arrivals, and ii) arrivals increment the counter and regenerate the phase. Concretely, for $k<K$:
\[
\Pr[(j,k)\to(j',k)] \text{ at rate } (\mathbf{T})_{jj'},
\]
\[
\Pr[(j,k)\to(j',k{+}1)] \text{ at rate } \mu_{j}\,\alpha_{j'}.
\]
At $k=K$, arrivals simply regenerate the phase without advancing the counter:
\[
\Pr[(j,K)\to(j',K)] \text{ at rate } (\mathbf{T})_{jj'} + \mu_{j}\,\alpha_{j'}.
\]
This implies the following generator:
\begin{equation}
\mathbf{A}=
\begin{pmatrix}
\mathbf{T} & \boldsymbol{\mu}\boldsymbol{\alpha}^\top & \mathbf{0} & \cdots & \mathbf{0}\\
\mathbf{0} & \mathbf{T} & \boldsymbol{\mu}\boldsymbol{\alpha}^\top & \ddots & \vdots\\
\vdots & \ddots & \ddots & \ddots & \mathbf{0}\\
\mathbf{0} & \cdots & \mathbf{0} & \mathbf{T} & \boldsymbol{\mu}\boldsymbol{\alpha}^\top\\
\mathbf{0} & \cdots & \cdots & \mathbf{0} & \mathbf{T} + \boldsymbol{\mu}\boldsymbol{\alpha}^\top
\end{pmatrix},
\label{eq:generator_ph}
\end{equation}
where each block is of dimension $d\times d$.

The traversal lasts exactly $\mtc$, thus the state distribution at the tail is $\mathbf{x}^{(\ell)}(\mtc) = \mathbf{x}^{(\ell)}\, e^{\mathbf{A} \mtc}$. Here, $\mathbf{P}^{\mtc} := e^{\mathbf{A} \mtc}$ acts as the transient transition matrix.

Next, we quantify the accumulated hits. Assigning to each state $(j,k)$ an instantaneous reward rate $r_{(j,k)}:=\mu_j$, the expected hits accrued over $\mtc$ follow directly from the Markov reward formula:
\begin{equation}
\mathbf{h} := \left(\int_0^{\mtc} e^{\mathbf{A} t}\,dt\right)\mathbf{r},
\label{eq:reward_ph}
\end{equation}
where $h_{(j,k)}$ is the expected number of hits collected during one traversal with head state $(j,k)$.

In the tail, the eviction hand executes the decrement rule in which counter-$0$ mass is purged and counters $1,\dots,K$ are shifted down by one. This operation can be encoded by $\mathbf{S}\in\mathbb{R}^{(K+1)d\times (K+1)d}$ utilizing $d\times d$ identity blocks on the sub-diagonal:
\[
\mathbf{S}=
\begin{pmatrix}
\mathbf{0} & \mathbf{0} & \mathbf{0} & \cdots & \mathbf{0}\\
\mathbf{I} & \mathbf{0} & \mathbf{0} & \ddots & \vdots\\
\mathbf{0} & \mathbf{I} & \mathbf{0} & \ddots & \mathbf{0}\\
\vdots & \ddots & \ddots & \ddots & \mathbf{0}\\
\mathbf{0} & \cdots & \mathbf{0} & \mathbf{I} & \mathbf{0}
\end{pmatrix}.
\]
Combining the in-traversal CTMC evolution with the discrete tail operation naturally yields an absorbing DTMC evaluated at the head of each traversal. In this chain, eviction corresponds to absorption, and the transient transition kernel is precisely 
$$\mathbf{Q} := \mathbf{P}^{\mtc}\mathbf{S}.$$
Furthermore, since mass only enters traversals at counters $\{0,\dots,K-1\}$, we restrict the state space to the head support $\mathcal{H} = \{(j,k): j\in\{1,\dots,d\},\ k\in\{0,\dots,K-1\}\}$ such that the transient kernel is the submatrix of $\mathbf{Q}$ restricted to $\mathcal{H}$:
\begin{equation}
\mathbf{Q}_H := \mathbf{Q}[\mathcal{H},\mathcal{H}]\in\mathbb{R}^{Kd\times Kd}.
\label{eq:Q_def_ph}
\end{equation}
Let $\mathbf{x}_H^{(0)}$ denote the restriction of the initial state $\mathbf{x}^{(0)}$ to $\mathcal{H}$. The accumulated head-state occupancy across all traversals is then evaluated via the fundamental matrix:
\[
\mathbf{v} = \sum_{\ell=0}^{\infty} \mathbf{x}_H^{(\ell)} = \mathbf{x}_H^{(0)}(\mathbf{I}-\mathbf{Q}_H)^{-1},
\]
and the expected number of traversals and total hits per admission follow as
\[
V = \mathbf{v}\,\mathbf{1},
\qquad
H = \mathbf{v}\,\mathbf{h}[\mathcal{H}].
\]
By \eqref{eq:pmiss_i} and \eqref{eq:occ_i},
\[
P_{\mathrm{miss}} = \frac{1}{1+H},
\qquad
C = \mtc\,\lambda\,P_{\mathrm{miss}}\,V.
\]
Finally, restoring the object index $i$, this construction yields $V_i$ and $H_i$. With PH arrival rate $\lambda_i = (-\boldsymbol{\alpha}_i^{\top}\mathbf{T}_i^{-1}\mathbf{1})^{-1}$, the per-object miss probability follows as~\eqref{eq:pmiss_i}. Global aggregations then follow as \eqref{eq:pmiss_agg} and \eqref{eq:tc_clockK}.

\subsection{Existence of Solution to~\eqref{eq:tc_clockK}}
\begin{proposition}
\label{prop:tc_existence_uniqueness}
For every $C\in(0,M)$ there exists a $\mtc\in(0,\infty)$ satisfying~\eqref{eq:tc_clockK}.
\end{proposition}

\begin{proof}
To see that $\mathcal{C}(\mtc) \in (0, M)$, we let $I_i(\mtc)$ denote the idle time between object $i$'s eviction and its subsequent admission. Since the expected cycle length between admissions is the sum of its residence time $\mtc V_i(\mtc)$ and its idle time $I_i(\mtc)$, the admission rate is natively evaluated to $\lambda_i P_{\mathrm{miss},i}(\mtc) = (\mtc V_i(\mtc) + \mathbb{E}[I_i])^{-1}$. Substituting this into~\eqref{eq:occ_i} gives:

\begin{align}\label{eq:Citauc}
     \mathcal{C}_i(\mtc) = \frac{\mtc V_i(\mtc)}{\mtc V_i(\mtc) + \mathbb{E}[I_i(\mtc)]} < 1 
\end{align}
provided that $\mtc V_i(\mtc) > 0$ and $\mathbb{E}[I_i(\mtc)] > 0$. The aggregate capacity across the full catalog therefore satisfies $\mathcal{C}(\mtc) \in (0, M)$.


The continuity of $\mathcal C(\mtc)$ follows directly from the PH matrix representation. For each object $i$, the transient transition matrix $e^{A_i \mtc}$ and the reward integral $\int_0^{\mtc} e^{A_i t} dt$ are continuous functions of $\mtc$. As the substochastic matrix $Q_{H,i}(\mtc)$ is transient for all finite $\mtc$, its spectral radius is strictly smaller than one, thus the fundamental matrix $(I - Q_{H,i}(\mtc))^{-1}$ is well-defined. Since matrix inversion is continuous on the set of non-singular matrices, $V_i(\mtc)$ and $H_i(\mtc)$ are continuous, therefore $P_{miss,i}(\mtc)$ and $\mathcal C_i(\mtc)$ are continuous and so too is $\mathcal C(\mtc)$.

It remains to verify the boundary limits. As $\mtc\to0$, the probability
of an arrival during a traversal tends to zero. Consequently,
$H_i(\mtc)\to0$ and $V_i(\mtc)\to1$, and \eqref{eq:Citauc} yields
\[
    \lim_{\mtc\to0}\mathcal C_i(\mtc)=0.
\]
As $\mtc\to\infty$, we have
$\mtc V_i(\mtc)\geq\mtc\to\infty$. Moreover, for a PH renewal process,
the expected residual time until the next arrival is uniformly bounded
over its finitely many phases. Hence
$\mathbb E[I_i(\mtc)]$ is uniformly bounded in $\mtc$, and
\eqref{eq:Citauc} gives
\[
    \lim_{\mtc\to\infty}\mathcal C_i(\mtc)=1.
\]
Therefore,
\[
    \lim_{\mtc\to0}\mathcal C(\mtc)=0,
    \qquad
    \lim_{\mtc\to\infty}\mathcal C(\mtc)=M.
\]
The intermediate value theorem now implies that, for every
$C\in(0,M)$, there exists at least one $\mtc\in(0,\infty)$ such that
$\mathcal C(\mtc)=C$.
\end{proof}

\section{Model Validation}
\label{sec:model-validation}

\begin{table*}[t]
  \centering
  \caption{Labels for experimental configurations.}
  \label{tab:label_map}
  \scriptsize
  \setlength{\tabcolsep}{4pt}
  \begin{tabular}{l|cccccc|cccccc|cccccc|cccccc}
    \toprule
    Workload & \multicolumn{6}{c|}{IRM} & \multicolumn{6}{c|}{Hyper-2} & \multicolumn{6}{c|}{Erlang-4} & \multicolumn{6}{c}{Hypo-4} \\ \midrule
    \textbf{Label} & \textbf{A} & \textbf{B} & \textbf{C} & \textbf{D} & \textbf{E} & \textbf{F} & \textbf{G} & \textbf{H} & \textbf{I} & \textbf{J} & \textbf{K} & \textbf{L} & \textbf{M} & \textbf{N} & \textbf{O} & \textbf{P} & \textbf{Q} & \textbf{R} & \textbf{S} & \textbf{T} & \textbf{U} & \textbf{V} & \textbf{W} & \textbf{X} \\
    $\theta$ & 0.5 & 1.1 & 0.5 & 1.1 & 0.5 & 1.1 & 0.5 & 1.1 & 0.5 & 1.1 & 0.5 & 1.1 & 0.5 & 1.1 & 0.5 & 1.1 & 0.5 & 1.1 & 0.5 & 1.1 & 0.5 & 1.1 & 0.5 & 1.1 \\
    $K$ & 0 & 0 & 1 & 1 & 15 & 15 & 0 & 0 & 1 & 1 & 15 & 15 & 0 & 0 & 1 & 1 & 15 & 15 & 0 & 0 & 1 & 1 & 15 & 15 \\
    \bottomrule
  \end{tabular}
\end{table*}

\begin{figure*}[h]
  \centering
  \includegraphics[width=2\columnwidth]{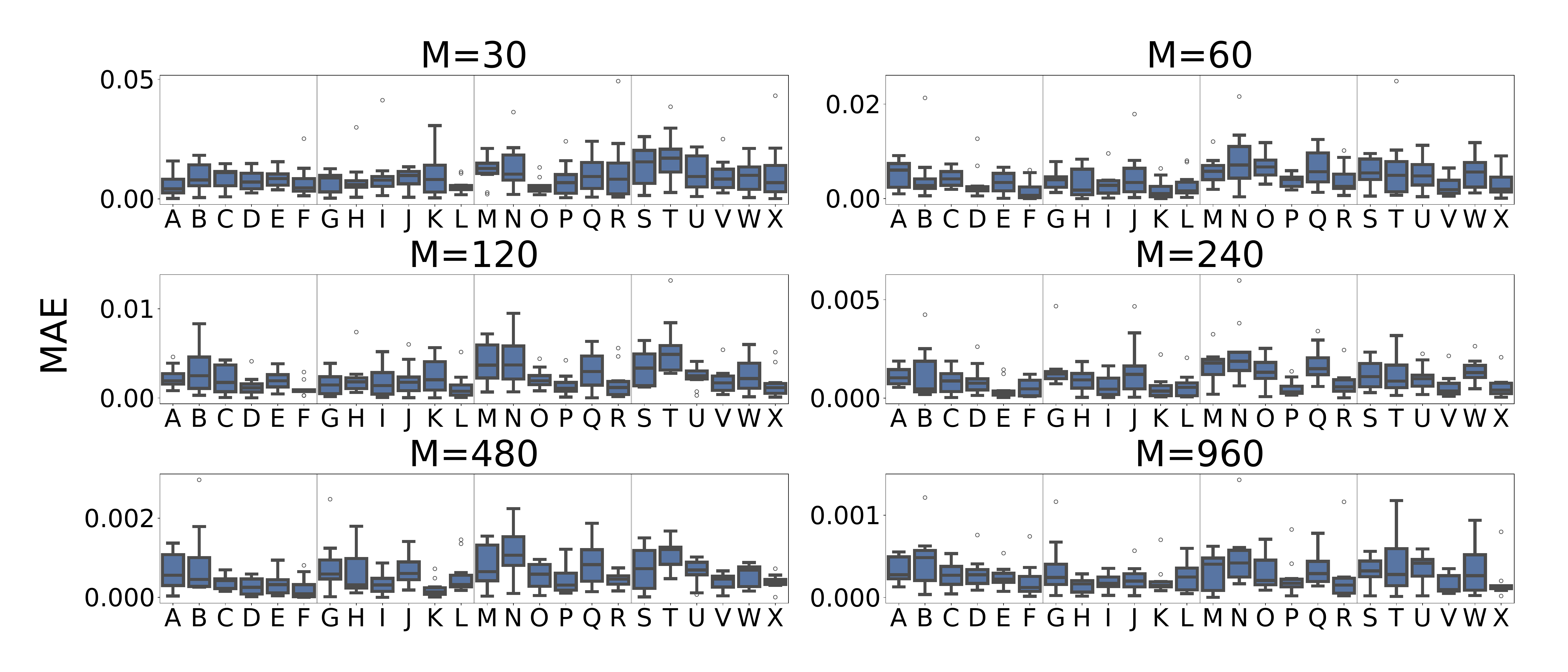}
  \caption{MAE between model-predicted and simulated miss ratio curves, computed over $\mtc\in[0,1.5M]$, across $M$ for the configurations in Table~\ref{tab:label_map}.}
  \label{fig:mae_boxes}
\end{figure*}
We validate the PH model of \S\ref{sec:clockK_ph} against
trace-driven simulations of the actual fixed-capacity
$\mathrm{CLOCK}(K)$ policy. For each value of $\mtc$ in a sweep over
$[0,1.5M]$, we evaluate \eqref{eq:tc_clockK} to obtain the corresponding
cache capacity $\mathcal C(\mtc)$ and compute the model-predicted miss
ratio. We then compare this prediction with the miss ratio obtained by
simulating $\mathrm{CLOCK}(K)$ at the corresponding cache capacity.
For each combination of
$M\in\{30,60,120,240,480,960\}$,
$\theta\in\{0.5,1.1\}$, and $K\in\{0,1,15\}$, accuracy is reported as
the mean absolute error (MAE) across the values of $\mtc$ in the sweep.
Each synthetic trace contains $10^7$ requests wherein object popularities follow a Zipf distribution:
\[
\lambda_i=\frac{i^{-\theta}}{\sum_{j=1}^M j^{-\theta}},\qquad i=1,\ldots,M.
\]
We evaluate four PH special cases spanning a range of coefficients of variation and constant, decreasing, and increasing inter-arrival hazard rates:
\begin{itemize}[leftmargin=*]
    \item \textbf{IRM.} A one-phase PH representation with $\boldsymbol\alpha_i^\top=[1]$ and $\mathbf T_i=[-\lambda_i]$.
    \item \textbf{Hyper-2.} A two-phase hyperexponential renewal process with
    \[
    \boldsymbol\alpha_i^\top=[1/2,1/2],\qquad
    \mathbf T_i=\lambda_i\begin{bmatrix}
    -\mu_1 & 0\\
    0 & -\mu_2
    \end{bmatrix},
    \]
    where $\mu_1/\mu_2=10$ and $1/(2\mu_1)+1/(2\mu_2)=1$.
    \item \textbf{Erlang-4.} A four-phase Erlang renewal process with
    \[
    \boldsymbol\alpha_i^\top=[1,0,0,0],\quad
    \mathbf T_i=\lambda_i
    \begin{bmatrix}
    -4 & 4 & 0 & 0\\
    0 & -4 & 4 & 0\\
    0 & 0 & -4 & 4\\
    0 & 0 & 0 & -4
    \end{bmatrix}.
    \]
    \item \textbf{Hypo-4.} A four-phase hypoexponential renewal process with
    \[
    \boldsymbol\alpha_i^\top=[1,0,0,0],\quad
    \mathbf T_i=\lambda_i
    \begin{bmatrix}
    -r_1 & r_1 & 0 & 0\\
    0 & -r_2 & r_2 & 0\\
    0 & 0 & -r_3 & r_3\\
    0 & 0 & 0 & -r_4
    \end{bmatrix},
    \]
    where
    \[
    (r_1,r_2,r_3,r_4)
    =(2s,4s,8s,16s),
    \quad
    s=\sum_{j=1}^4\frac{1}{2^j}.
    \]
\end{itemize}
Figure~\ref{fig:mae_boxes} shows the MAE distributions for each parameter combination labeled in Table~\ref{tab:label_map}. The predictions closely track simulation across all four arrival models, catalog sizes, skew parameters, and counter caps, with errors generally decreasing as the catalog size grows.

\section{Schur-Concavity under IRM}
\label{sec:clockk_properties}

\begin{figure}[h]
    \centering
    \includegraphics[width=0.6\columnwidth]{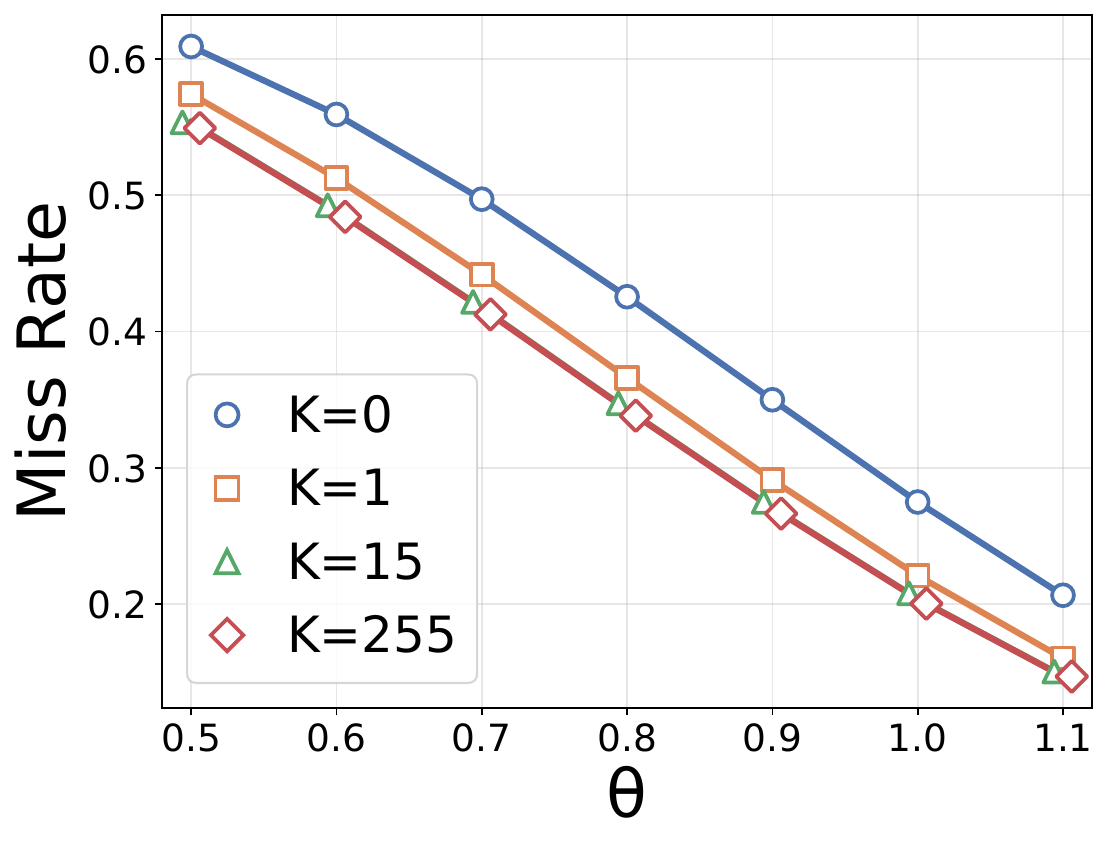}
    \caption{CLOCK($K$) miss ratio under IRM with Zipf popularity $\lambda_i \propto i^{-\theta}$ ($M=10^3$, $N=10^7$). 
    Cache size $C=300$; $\theta \in \{0.5,0.6,0.7,0.8,0.9,1.0,1.1\}$. Miss ratio decreases with $\theta$ and is non-increasing in $K$, with marginal gains beyond $K=15$.}
    \label{fig:hr_irm}
  \end{figure}
{This section reports empirical evidence for the Schur-concavity} of the CLOCK($K$) miss ratio with respect to object popularity under IRM.

We begin with below definitions.
\begin{definition}[Majorization]
Vector $\boldsymbol X=(X_1,\dots,X_k)$ is said to majorize vector $\boldsymbol Y=(Y_1,\dots,Y_k)$ (written $\boldsymbol Y\prec\boldsymbol X$) iff
\[
\sum_{i=1}^j \hat Y_i\le \sum_{i=1}^j \hat X_i,\qquad j=1,\dots,k-1,
\]
\[
\sum_{i=1}^k \hat Y_i=\sum_{i=1}^k \hat X_i,
\]
where the notation $\hat X_i$ (resp. $\hat Y_i$) is taken to be the $i$th largest element of $\boldsymbol X$ (resp. $\boldsymbol Y$).
\end{definition}
\begin{definition}[Schur-convexity / Schur-concavity]
A real-valued symmetric function $f$ is Schur-convex if
\[
\boldsymbol x\prec\boldsymbol y\ \Rightarrow\ f(\boldsymbol x)\le f(\boldsymbol y),\qquad \forall\,\boldsymbol x,\boldsymbol y.
\]
It is Schur-concave if $\boldsymbol x\prec\boldsymbol y\Rightarrow f(\boldsymbol x)\ge f(\boldsymbol y)$.
\end{definition}
We also find that the following lemma~\cite[C.1]{marshall1979inequalities} is useful.
\begin{lemma}
\label{lem:marshall1}
If $I \subset \mathbb{R}$ is an interval and $g : I \to \mathbb{R}$ is convex, then
\[
\phi(x) = \sum_{i=1}^{n} g(x_i)
\]
is Schur-convex on $I^n$. Consequently, $x \prec y$ on $I^n$ implies $\phi(x) \le \phi(y)$.
\end{lemma}


As shown in Figure~\ref{fig:hr_irm}, our trace-driven simulations demonstrate that under IRM, CLOCK($K$) miss ratio decreases as the popularity skew of the object increases. Since a larger Zipf parameter $\theta$ majorizes a smaller one, Schur-concavity holds for CLOCK($K$) across $K \in \{0, 1, 15, 255\}$ and $\theta \in [0.5, 1.1]$.
While this appears to contradict~\cite[p.~511, Lemma~2]{van1993properties} in which FIFO is proven Schur-convex over object popularity; a close examination of their proof reveals a minor sign error that reverses the stated property. {This is easily verified for} $C=1$ where $P_{\mathrm{miss}} = 1 - \sum_i \lambda_i^2$ is Schur-concave by Lemma~\ref{lem:marshall1}. A detailed correction is provided in Appendix~\ref{sec:fifo_note}.

\section{Conclusion}
We derive a TTL approximation for the steady-state miss ratio of CLOCK($K$) under phase-type renewal traffic and validate it against trace-driven simulations. We also provide empirical evidence that the resulting miss ratio is Schur-concave in the object-popularity vector under IRM.

A promising avenue for future work is to formally establish the structural connection between CLOCK($K$) and Ran-CLOCK($K$)~\cite{ran-clock} under the IRM in the $K \to \infty$ asymptotic limit.

\newpage

\appendix
\section{A Remark on Prior Work}
\label{sec:fifo_note}
Here we give a remark on a prior proof of FIFO Schur-convexity in~\cite{van1993properties}. We begin with the following lemma~\cite[Sec.3 A.2.b]{marshall1979inequalities}:
\begin{lemma}[Marshall--Olkin]
\label{lem:clockk_mo}
Let $\mathcal A$ be a set with the property
\[
\boldsymbol y\in\mathcal A\ \text{and}\ \boldsymbol x\prec\boldsymbol y\ \ \text{implies}\ \boldsymbol x\in\mathcal A.
\]
A continuous function $\phi$ defined on $\mathcal A$ is Schur-convex on $\mathcal A$ if and only if $\phi$ is symmetric and
\[
\phi(x_1,\,s-x_1,\,x_3,\dots,\,x_n)\ \text{is increasing in}\ x_1\ge s/2
\]
for each fixed $s,\,x_3,\dots,\,x_n$.
\end{lemma}

In~\cite[Theorem~2(b)]{van1993properties}, the FIFO miss ratio is proven Schur-convex in the reference-probability vector $\boldsymbol p$ by invoking a reformulation of Lemma~\ref{lem:clockk_mo}:
\begin{lemma}[van den Berg and Towsley]
A function $f$ is Schur-convex iff $f$ is symmetric and 
$$
f(\lambda q, (1-\lambda)q, y_3, y_4, \cdots, y_k)
$$
is a non-decreasing function of $\lambda$ for $\lambda\in(0,1/2]$.
\end{lemma}
However, this invocation applies the Marshall--Olkin slice test on the wrong side. In the original formulation, the monotonicity is checked for the larger coordinate ($x_1\ge s/2$), whereas $\lambda\in(0,1/2]$ corresponds to $p_1=\lambda q\le q/2$. The correct range is $\lambda\in[1/2,1)$ (or equivalently, swap the roles of $p_1$ and $p_2$):
\begin{lemma}[van den Berg and Towsley corrected]
    A function $f$ is Schur-convex iff $f$ is symmetric and
    $$f(\lambda q,(1-\lambda)q,y_3,y_4,\cdots,y_k)$$ is a non-decreasing function of $\lambda$ for $\lambda\in[1/2,1)$.
\end{lemma}

\bibliographystyle{plain}
\bibliography{sample-base}

\end{document}